\documentstyle[aas2pp4,psfig,rotate]{article}
 
\newcommand{\kms}{$\,{\rm km\,s^{\scriptscriptstyle -1}}$}
\newcommand{\gtsim}{\ {\raise-0.5ex\hbox{$\buildrel>\over\sim$}}\ }
\newcommand{\ltsim}{\ {\raise-0.5ex\hbox{$\buildrel<\over\sim$}}\ }

\journalid{}{}
\articleid{11}{14}  

\begin{document}

\title{HST Imaging of Globular Clusters in the Edge--on Spiral Galaxies
NGC 4565 and NGC 5907}

\author{Markus Kissler-Patig \altaffilmark{1}}
\affil{European Southern Observatory, Karl-Schwarzschild-Str.~2, D-85748
Garching, Germany \\ email: mkissler@eso.org}
\affil{University of California Observatories / Lick Observatory, 
University of California, Santa Cruz, CA 95064}

\author{Keith M. Ashman}
\affil{Department of Physics and Astronomy, University of Kansas,
	Lawrence, KS 66045 \\ e-mail: ashman@kusmos.phsx.ukans.edu}

\author{Stephen E. Zepf}
\affil{Department of Astronomy, Yale University, New Haven, CT 06520
\\ e-mail: zepf@astro.yale.edu}

\author{Kenneth C. Freeman}
\affil{Mount Stromlo and Siding Spring Observatories, ANU, Private Bag, 
Weston Creek PO, 2611 Canberra, ACT, Australia \\ e-mail: kcf@mso.anu.edu.au}

\altaffiltext{1}{Feodor Lynen Fellow of the Alexander von Humboldt Foundation}

\begin{abstract}

We present a study of the globular cluster systems of
two edge--on spiral galaxies,
NGC~4565 and NGC~5907, from WFPC2 images in the F450W and F814W filters.
The globular cluster systems of both galaxies appear to be similar
to the Galactic globular
cluster system. In particular, we derive total numbers of globular
clusters of $N_{GC}(4565)= 204\pm38 ^{+87}_{-53}$ and $N_{GC}(5907)=170\pm41 
^{+47}_{-72}$ (where the first are statistical, the second potential systematic
errors) for NGC~4565 and NGC~5907, respectively.
This determination is based on a comparison to the Milky Way system,
for which we adopt a total number of globular clusters of $180\pm20$.
The specific frequency of both
galaxies is $S_{N}\simeq0.6$: indistinguishable from the value for the 
Milky Way. The similarity in the globular cluster systems of the two
galaxies is noteworthy since they have significantly 
different thick disks and bulge-to-disk ratios. This would suggest that
these two components do not play a major role in the building up of a
globular cluster system around late--type galaxies. 

\end{abstract}

\keywords{galaxies: individual (NGC~4565, NGC~5907)
--- galaxies:formation --- galaxies: star clusters}

\section{Introduction}

Globular clusters in extragalactic systems have established themselves
as powerful diagnostics of the star formation history, dynamics, and
structure of their host galaxies (see Ashman \& Zepf 1998 for a recent
review). Most of the observational studies of extragalactic
globular cluster systems (GCSs) have concentrated on
the systems of early--type galaxies, where globular
clusters are easily identified against the smooth, dust free background. 
These studies have uncovered the characteristic properties of such
globular cluster systems, as well as notable galaxy-to-galaxy variations
in these properties. In comparison, information on the globular cluster systems
of late-type galaxies is sparse. 

The use of globular cluster systems as probes of the formation and 
evolution of galaxies is severely limited by this scarcity of data for
spiral galaxy GCSs. There are two primary problems. The first is that
much of the interpretation of early-type galaxy GCSs has used the well-studied
globular cluster system of the Milky Way as a benchmark. For example, the
result that the globular clusters of elliptical galaxies are more metal-rich
in the mean than those of spiral galaxies is largely based on the mean
metallicity of Milky Way globulars. Consequently, it is extremely important
to establish whether the properties of the Milky Way globular cluster system
are typical of spiral galaxy GCSs.

The second issue is that a better understanding of the GCSs
of late--type galaxies is needed in order to fully utilize GCSs as probes
of the formation and early evolution of galaxies. This is self-evident
for the case of late--type galaxies, but is also a critical issue
for constraining models of the formation and early evolution of elliptical
galaxies. This is particularly important for testing the
predictions of the merger model, in which
elliptical galaxies are formed in the merger of spiral galaxies.
In this picture, the GCSs of elliptical galaxies are composite systems. 
One population of clusters is associated with the progenitor spirals,
while a second population of clusters forms in the merger event.
Ashman and Zepf (1992) described several testable predictions arising from this
scenario. However, the limited information on the the GCSs of spirals
leads to uncertainties in these predictions since it is currently unclear
what constitutes a typical spiral galaxy GCSs. One particular problem
is that the characteristic specific frequency of globular clusters 
around spirals is poorly known. The absence of firm constraints on
the number of globular clusters contributed by the progenitor spirals to
an elliptical galaxy GCS leads to uncertainties in many of the detailed
predictions for these systems that follow from the merger model. One of the
key results of the present study is to add to the database of globular
cluster specific frequencies for spiral galaxies.

In this paper, we present an imaging study
 of the globular cluster systems of two
edge--on spiral galaxies: NGC 4565 and NGC 5907. Globular clusters were
already statistically detected in NGC 4565 as stellar-like objects
above the mean background by van den Bergh \& Harris (1982) and 
Fleming et al.~(1995). The former detected $\sim$ 100 clusters to an
equivalent limiting magnitude of $B\simeq23.3$, and found the clusters
to be compatible with a surface density profile of the form $\sigma \propto 
r^{-2.5}$, similar to the profile of the Milky Way system.
The latter study determined the peak of the globular cluster luminosity 
function to be at V$_0=22.63\pm0.21$. Fleming et al.~(1995) extrapolated
the observed number of clusters ($71\pm30$) to a total
number of $180\pm45$ globular clusters and a 
specific frequency of $S_N=0.43\pm0.11$ under their assumptions for
luminosity and distance. NGC 5907 has been searched for globular clusters by
Harris, Bothun \& Hesser (1988), but no globular cluster could be
detected. The authors speculated that the lack of globular clusters was
related to the non--existence of a central bulge.

In Section 2, we present
the properties of the 
 galaxies and the observations, and describe the photometry against a
highly variable background. In Section 3, the results are analyzed. In
Section 4 we compare the globular cluster systems to each other and to
the Milky Way. Concluding remarks are made in Section 5.

\section{Observations and Data Reduction}

\subsection{Observations}

\subsubsection{The target galaxies}

We obtained WFPC2 observations of the nearby, edge--on
spiral galaxies, NGC~4565 and NGC~5907 (Program GO 6092). 
Basic properties of the two galaxies are given in Table 1 and a composite
of our images is shown in Fig.~1.
NGC~4565 is classified as an Sb galaxy (RC3, Hubble Atlas) with
an inclination of $86.5^{\circ}$ (Garc\'{\i}a-Burillo et al.~1997), and is at  
a distance of 10 Mpc, based on
surface brightness fluctuations and planetary nebulae
(Jacoby et al.\ 1996 and references therein).
NGC~5907 is classified as an Sc galaxy (RC3, Hubble Atlas),
with an inclination of $87^{\circ}$ (Morrison et al.\ 1994), and 
is at a distance of 12 Mpc, based on the Tully--Fisher relation
(Bottinelli et al.~1988, Sch\"oniger \& Sofue 1994).

The two galaxies are very similar to each other in luminosity, but
differ considerably in their disk characteristics.
NGC~4565 has a prominent thick disk, accounting for
roughly $5\%$ of the total luminosity (van der Kruit \& Searle 1981).
NGC~5907 has a less prominent bulge and an upper limit to the thick
disk luminosity of 0.6\% of the thin disk value---a relative contribution
of less than $10\%$ of the thick disk of the Milky Way (Morrison et al.~1994).

\subsubsection{The observations}

For both of our target galaxies, NGC~4565 and NGC~5907,
we obtained two pointings with the WFPC2 on HST. As shown
in Fig.~1, we aligned each pointing so that the major
axis was parallel to the edges of two WF chips, the third WF chip and
the PC sampling part of the halo.

For every pointing, images in the F450W and F814W filters were obtained
with total integration times ranging from 600 sec to 780 sec in the F450W 
filter, and of 480 sec in the F814W filter. Every pointing was split in
three exposures shifted by 0.5\arcsec.
	
\subsection{Data Reduction}

\subsubsection{Basic reduction and photometry}

The basic image reduction was carried out under {\tt IRAF}.
For every pointing, the three calibrated science images provided by the Space 
Telescope Science Institute were shifted by 0.5\arcsec\ (using the task
{\tt imshift}) and trimmed in order to be aligned. They were then combined 
with the task {\tt crrej} that rejected the vast majority of cosmic
rays.The images were transformed into {\tt FITS}
format for the photometry.

The photometry was carried out using the {\tt SExtractor} software
(Bertin \& Arnouts 1996). The finding parameters were set to 2 connected
pixels 2.5$\sigma$ above the local background (computed in a 10 by 10
pixel mesh). The computation of a local background in a relatively fine
grid was chosen to allow the finding criteria to be equivalent in the
various parts of the image, in particular towards the disks where the
background becomes very irregular. Typically, down to an equivalent B
magnitude of 25.5, 0 to 2 objects were
detected in both the F450W and F814W image on the PC, 8 to 18 on the WF
``halo'' field, and 130 to 340 on the WF fields including disc and
bulge. The latter included a large number of stellar associations and H{\tt
II} regions.
Positions on the sky (RA and DEC) were computed for all objects using
the task {\tt metric} in the {\tt STSDAS} package. Distance from the
center, along the disc, and orthogonal distance from the disc
were computed using position angles of $135.5^\circ$ for NGC 4565 
(Rupen 1991) and $156^\circ$ for NGC 5907 (Garc\'{\i}a-Burillo et 
al.~1997). 

The photometry was carried out with 2 pixel radius apertures in
order to avoid being affected by background irregularities.
Corrections to get the equivalent magnitude to a 0.5\arcsec\ aperture were
determined from isolated bright objects on the images and found to
be 0.35 and 0.18 mag for the PC and WF F450W measurements
respectively, and 0.55 and 0.21 mag in the F814W filter, with errors of
$<10$\%. 
Additional 0.1 magnitudes were added to obtain the total magnitude of the 
objects (derived for the WFPC2 point spread functions, following Holtzman et 
al.~1995). Note that at the distance of our galaxies the largest globular 
clusters will be resolved and these corrections underestimated. This will
hardly affect the colors since the luminosity profiles in F450W and
F814W are similar,
but could significantly affect the individual magnitudes, as
a function of the object  size (cf.~Kundu \& Whitmore 1998, Puzia et al.~1999).
Finally, the calibration relation given in Holtzman et al.~(1995) were used to 
obtain Johnson--Cousins B and I magnitudes.

\subsubsection{Globular cluster selection}

Extensive artificial star experiments with artificial point sources and 
artificial globular clusters were carried out. For the latter WFPC2 point
spread functions obtained with {\tt Tiny Tim} (Krist \& Hook 1997) and
convolved with a Modified Hubble law of core radius 0.1\arcsec\ and 0.2\arcsec\
were used 
\footnote {The core radius of the modified Hubble law is
almost identical to the ``core'' radius of a King profile, the King
radius being the radius where the projected density of an isothermal
sphere drops to almost half (0.5013) as oppose to half (0.5) for the
core radius of the Modified Hubble law.}. The latter correspond to core
radii of about 4 to 10 pc at the distance of 10 to 12 Mpc, and lie in the upper
range of the values measured in the Milky Way (only 27\% of the Milky Way
cluster have a core radius larger than 4 pc in the compilation of Harris
1996).
The resulting FWHM of our artificial objects range from 0.8 to 1.5 pixels and
1.5 to 2.5 pixels for artificial stars and globular clusters
respectively, at B$<23.5$. These allowed us to define robust
selection criteria for globular clusters based on the FWHM of the
detected objects ($>0.7$ and $<3.0$ pixels, as returned by {\tt SExtractor}).
These selection criteria were applied 
down to B$=23.5$; at fainter magnitudes the signal--to--noise was not 
sufficient to discriminate with high confidence between point sources and 
extended objects. 
Note that down to this magnitude limit our finding algorithm is 100\% complete 
and objects will only be missed if they are physically obscured, e.g.~by dust.

We further introduced a cut in color and magnitude. For the color cut,
we chose B--I $>1.2$, roughly 0.2 magnitudes bluer than the bluest
globular clusters in the Milky Way (see also Fig.~3). No upper limit was chosen
to avoid excluding reddened objects. As a magnitude range, we chose
B$>20.0$, corresponding to B$>-10.0$ and $-10.4$ at the respective
distances of NGC~4565 and NGC~5907. 
As a reference, the brightest clusters in the
Milky Way and M31 have B$=-9.4$ ($\omega$ Cen) and B$=-9.7$ (Mayall II).
Our upper magnitude cut was dictated by the FWHM selection, 
only reliable down to B$=23.5$ which was used as the limiting magnitude.
The peak of the globular cluster luminosity function is expected at 
B$\simeq -7.1$, as derived by Della Valle et al.~(1998), using the new
{\tt HIPPARCOS} distance calibration and the list of Milky Way clusters of
Harris (1996). This corresponds to B$\simeq22.9$ and B$\simeq23.3$ at
the distance of the two galaxies, so that we pass the peak of the
luminosity function in both cases.

The expected number of foreground stars contaminating our images and
passing our selection criteria were estimated from galactic models
(Ibata, priv.~com.) and the Hubble Deep Field to be $1\pm1$ per pointing.
The number of contaminating galaxies was estimated from several
fields of the Medium Deep Survey (Griffiths et al.~1993) and our
selection criteria turned out to be robust against background galaxy
contamination at these bright magnitudes (B$<23.5$) estimated to be
$0\pm1$ per pointing.

The final samples include 40 and 25 globular clusters 
in NGC~4565 and NGC ~5907 respectively (see Table 2 and 3). 
All candidates (except one in NGC 5907 identified as a star) were confirmed
by a visual inspection. The distribution of globular clusters, together
with all objects detected is shown in Fig.~2. The colors of the 
globular cluster candidates in NGC~4565, as well as visual inspection of
the galaxy dust lane, suggests that the majority of the clusters with
positive Z suffer significant extinction. The identifiable regions of dust
are less extended in NGC~5907 and there is less direct evidence for
significant extinction, but we cannot rule out reddening of the clusters
in this galaxy.

Color magnitude diagrams for the detected globular clusters in NGC 4565
and NGC 5907 are shown in Fig.~3, together with one for Milky Way
globular clusters (uncorrected for reddening) as comparison. The mean
colors and dispersion for the samples are $(B\!-\!I)=1.9\pm0.4$ in NGC 4565, 
$(B\!-\!I)=1.7\pm0.3$ in NGC 5907, and $(B\!-\!I)=2.0\pm0.4$ in the Milky Way 
($(B\!-\!I)=1.8\pm0.3$ when corrected for reddening, corresponding to a mean
metallicity of [Fe/H]$=-1.3\pm0.5$ dex). This tends to support the evidence
that reddening is more of an issue in NGC~4565 than NGC~5907, at least
if the {\it intrinsic} colors ($\sim$ metallicities) of the GCSs of the three galaxies
are similar. One cautionary note in this regard is that the bluer mean
colors of the NGC~5907 clusters is partly driven by some very blue objects
amongst the globular cluster candidates. Since these blue objects are
not preferentially located near the disk, it is unlikely that they
are young compact star clusters, but it is possible that one or two such objects
maybe be included in the sample.
Despite these complications, the small differences ($\simeq 0.1$) in the mean
$(B\!-\!I)$ values probably 
suggest only moderate differences in the mean metallicities
($<0.4$ dex following the conversion of Couture et al.~1990).

\section{Total Numbers of Globular Clusters and Specific Frequencies}

\subsection{The total number of globular clusters}

In order to estimate the total number of globular clusters around
NGC~4565 and NGC~5907, three extrapolations are required. The first
is to account for
globular clusters fainter than our magnitude limit. The second is
for globular clusters that are not covered by our imaging field. The
third is for globular clusters lying behind the disk and bulge and
being obscured. 

\subsubsection{The approach: a direct comparison with the Milky Way}

We chose to estimate the counts by direct comparison with the Milky 
Way and estimate the total number of globular clusters in the two galaxies
by comparing our observations with the same spatial region and 
equivalent magnitude limit in our Galaxy. 
This approach has the advantage of taking into account all three
extrapolations at once without propagating unnecessary errors. 
It is motivated by the small number of objects detected and the limited spatial
coverage of our observations. These do not allow us to
derive the spatial profile of either globular cluster system, necessary
to estimate the spatial extrapolation. Further, we would still need a
model to account for the clusters behind the disk.

Specifically, we created
a mask for each galaxy based on the areal coverage of the two WFPC2
pointings of our observations. These masks excluded regions between
chips and the edges of chips where the images were trimmed. Also excluded
were small regions ($\sim$ 150 $\Box$\arcsec) centered on the bulges where our
artificial star experiments indicated the data to be incomplete. 
By imposing the relevant mask onto
the Milky Way clusters and using an absolute magnitude cut appropriate
to $B=23.5$ at the distances of the two galaxies, we determined the 
number of Milky Way globular clusters that would have been detected with our
observations.  
We used positions and magnitudes of Milky Way
globular clusters taken from the McMaster catalog (Harris 1996) to create a 
2-dimensional spatial distribution by projecting Galactocentric $(X,Y,Z)$
coordinates into the $Y-Z$ plane. 
Figures 4 and 5 give a visual representation of the results of this process.  

For each galaxy, there are four possible orientations of the mask that
preserve the alignment of the pointings along the major axis of the
galaxy. We therefore repeated the above exercise for all four orientations
and in each case counted the number of Milky Way globular clusters that
were ``detected'' (see Table 4). The mean number of clusters from these four 
possibilities was used to estimate the total number of globular clusters, 
$N_{GC}$, in each galaxy, using the expression:
$$
N_{GC} = N_{GC}({\rm Milky Way}) \frac{ N_{obs} }{ N_{mask} }
$$
where $N_{GC}({\rm Milky Way})$ is the total number of globular
clusters in the Milky Way ($180\pm20$, Ashman \& Zepf 1998), $N_{obs}$ is the 
number of globular clusters observed in our target galaxies (40 and 25
for NGC 4565 and NGC 5907, respectively, corrected to 38 and 23 when
accounting for fore-/background contamination), and $N_{mask}$ is
the mean number of globular clusters from the four mask orientations
(see Table 4).

The method relies on several assumptions that we discuss in turn. 

\subsubsection{The Milky Way globular cluster sample}

The McMaster catalog lists 141 globular clusters, 134 of which
have Galactocentric coordinates, and 112 of which have de-redenned
absolute B magnitudes (which served for the magnitude cut). 
However, the clusters
with no reported $B$ magnitudes are almost exclusively more than one
magnitude fainter than the peak of the luminosity function (based on 
their available absolute $V$ magnitudes) and would also be undetected at the 
distances of our target galaxies. 

We assumed the total number of globular clusters in the Milky Way to be 
$180\pm20$. The majority of undetected Milky Way globulars are assumed to be 
behind the Galactic bulge. 
The counterparts of such globular clusters in NGC~4565 and
NGC~5907 would also be undetected in our observations due to
obscuration by disk and bulge. 
The scale length and scale heights of the thin disk of all
three galaxies is not too different (e.g.~van der Kruit \&  Searle 1981, 
Morrison et al.~1994, Kent et al.~1991). We therefore assume
than the number of obscured objects is the same in all galaxies (within
20\%).

Thus while our Milky Way sample is not complete, the missing
clusters tend to be ones that we would not detect in NGC~4565 and NGC~5907
anyway.

\subsubsection{Extrapolation over the luminosity function}

We assumed that the globular cluster
luminosity functions of NGC 4565 and NGC 5907 are similar to the one of
the Milky Way, i.e.~roughly Gaussian with a peak at B$=-7.1\pm0.1$ and a 
standard deviation of $\sigma = 1.1$ (see Sect.~2.2). This assumption of a
`universality' of the globular cluster luminosity functions is well
supported by all recent observations (e.g.~Whitmore 1997 for a recent
review). For our assumed distances (see Table 1), our limiting magnitude of 
B$=23.5$ corresponds to $0.6\pm0.1$ and $0.2\pm0.1$ magnitudes past the
peak for NGC 4565 and NGC 5907 respectively. This corresponds to $0.54\pm0.09 
\sigma$ and $0.18\pm0.09 \sigma$ past the peak. In other words, we
sampled $70\pm3$\% and $57\pm3$\% of the luminosity functions
respectively, and had to implicitly extrapolate over the remaining fraction. 
This extrapolating was taken care of by a corresponding cut in the
absolute magnitude in the Milky Way sample.
We stress that down to the magnitude limit, we do not suffer from finding
incompleteness due to the software finding algorithm.

\subsubsection{Extrapolation over the spatial distribution}

Another caveat is the implicit assumption that
the spatial profile of the Milky Way globular cluster system is similar
to that of the globular cluster systems of NGC~4565 and NGC~5907. For
NGC~4565, the available photographic data are consistent
with this assumption (see Sect.~1). For NGC~5907,
we assume it also to be the case (but see Sect.~3.3). 
If the spatial profile of the globular cluster system of NGC~5907 was
markedly shallower than that of the Milky Way, we would detect a higher
fraction of globular clusters in the ``halo'' WF and PC fields than we
actually do. On the other hand, if the spatial profile was much steeper,
it would not greatly affect our estimate of $N_{GC}$ since our areal
coverage includes roughly half of all the globular clusters for 
density profiles comparable to that of the Milky Way.

\subsubsection{Errors in the distance}

Errors in the distances to the galaxies lead to 
uncertainty in both the limiting magnitude and
size of the mask used to derive the total number of clusters around 
each galaxy. The distance to
NGC~4565 is fairly well established, with three independent secondary
indicators all agreeing closely with a value of 10~Mpc. The distance
to NGC~5907 was derived from the Tully--Fisher relation and found to be
12 Mpc by independent groups. To estimate the effects of distance variations
on our derived values, we carried out the same procedure as above with
the galaxies at assumed distances of $\pm 20$~\% their preferred values,
affecting both mask size and limiting magnitude.
The results of this exercise are summarized in Table 4. 

\subsubsection{The total numbers of globular clusters}

In summary, the above considerations lead us to total numbers of
globular clusters for the two galaxies of $N_{GC}(4565)=204\pm38$ and 
$N_{GC}(5907)=170\pm41$, following the equation of Sect.~3.1.1, where
$N_{obs}$ was corrected for fore-/background contamination. 
The error is the statistical error only: it includes Poisson errors in the 
observed number of globular clusters, Poisson errors in the average number of
Milky Way globular clusters in the mask (derived from the 4 different
orientations), and errors in the number of contaminating objects.
To this error, potential systematic errors should be added.
First, we estimate up to 20\% difference in the number
of obscured clusters between the Milky Way and our target galaxies. 
Second, we assumed for the Milky Way a total number of clusters of $180\pm20$; 
for a different assumption, our counts should be adjusted accordingly. 
Note that this systematic error does not influence any direct comparison 
between our two galaxies. The third systematic error comes from our assumed
distances. For a 20\% error in distance, our results would vary by
$^{+74}_{-29}$ for NGC 4565 and $^{+29}_{-62}$ for NGC 5907,
mainly due to the different limiting magnitude.
Note that this systematic error significantly affects the total numbers
of clusters, but has a much smaller effect on the derived
specific frequencies (see below) due to the varying total magnitude
of the galaxy which compensates somewhat the varying total number of
clusters.
Finally, for NGC 5907 an additional uncertainty is present due to the 
lack of knowledge on the exact shape of the globular cluster density profile.

Thus, formally, the total number of globular clusters are $N_{GC}(4565)=
204\pm38 ^{+87}_{-53}$ and $N_{GC}(5907)=170\pm41 ^{+47}_{-72}$, where
the first error is the statistical one, the second error the systematic one.

For NGC 4565, our result is in good agreement with that of Fleming et
al.~(1995), who followed a completely different approach. This gives us
confidence that our approach is also valid for NGC 5907 and the previous 
non--detection of globular cluster should probably be attributed to the
poor observing conditions.

\subsection{The specific frequencies}

One can also express these numbers in terms of a specific frequency
$S_N=N_{GC}10^{0.4(M_V+15)}$, where $M_V$ is the absolute visual
magnitude of the parent galaxy. Using the absolute magnitudes 
of the two galaxies given in Table 1 and the total numbers from the
previous section, we obtain $S_N=0.56\pm0.15$ and $S_N=0.56\pm0.17$ for
NGC 4565 and NGC 5907, respectively. The
errors include the random errors from above, as well as an error of 0.2
magnitudes in the de-reddened apparent magnitudes of the galaxies. 
The effect of the systematic errors would be to lower the $S_N$ values
by 0.05, should the total number of globular clusters in the Milky Way
be $160\pm20$; and to vary $S_N$ by less than 0.1 should the distance of the
galaxy vary by 20\% (the increase/decrease in total number being
partly compensated by an increase/decrease in the total magnitude of the
galaxy). 

The quantity $S_N$ was introduced by
Harris \& van den Bergh (1981) primarily for use in elliptical galaxies
where there is little variation in the stellar population from one
galaxy to another. 
Zepf \& Ashman (1993) attempted to account for stellar
mass-to-light ratio variations in a {\it statistical} sense by
defining a parameter $T$ to be the number of globular clusters per unit
($10^9 M_{\odot}$) stellar mass of a galaxy. Conversion from galactic
luminosity to stellar mass was achieved by assuming a characteristic
$M/L_V$ for each morphological type of galaxy. This is not an ideal
procedure for individual galaxies, but it does tend to minimize
differences in $S_N$ generated by stellar population differences.
Thus the approach allows us to obtain
a comparison of the number of globular clusters in our target galaxies
with other late--type galaxies studied to date. For the conversion from
luminosity to mass, we follow Zepf \& Ashman (1993) and use $M/L_V$
values of 5.4, 6.1 and 5.0 for Sa, Sab-Sb and Sbc-Sc galaxies, respectively 
(cf.~Faber \& Gallagher 1979).
The $T$ value for NGC 4565 is $1.0\pm0.3$ if it is assumed to be of type Sb and
1.2 if its type is Sbc (errors are of the order 20\%, see above). For NGC 5097 
(type Sc) $T=1.3\pm0.3$. We comment on these values below.

\subsection{$Z$-distributions}

A visual comparison of the detected globular clusters in NGC~4565 and NGC~5907 
with the masked Milky Way distributions (Fig.~4 and 5) suggests that
the globular cluster system of NGC~5907 may be more flattened than
that of the Milky Way.   It is difficult to derive shape information
directly for the globular cluster system of this galaxy from our small
sample of 25 objects, as well as the asymmetric geometry of the field
of view. A comparison of the $Z$-distributions of NGC~5907 with that of the
masked Milky Way dataset suggests a difference at a 2-$\sigma$ level
according to a K-S test (NGC~4565 and the Milky Way show no detectable
difference). The inner regions of the Milky Way globular cluster system
is itself known to be somewhat flattened, primarily because of the presence
of a thick disk population of globular clusters (e.g.\ Zinn 1985)
so our data are
consistent with a flattened globular cluster distribution in NGC~5907, but
do not demand it. The previous non--detection of globular clusters in
NGC 5907 (Harris et al.~1988) could also point towards a deficiency of
objects in the halo, although we can only speculate on this point. Their
ground based study was based on object over-densities, and was probably
unable to detect clusters close to the disk. With a limiting
magnitude similar to ours, the density of globular clusters
in the halo would be only marginally detectable in a statistical comparison
with a background field.

The shape of the cluster system in NGC 5907 is of some interest since it has 
been reported by Sackett et
al.~(1994) that this galaxy has a highly flattened halo ($\epsilon\sim0.4$) 
with a flat density profile falling off roughly as $x^{-2.2}$, compared to
$x^{-3.5}$ for the Milky Way halo and halo clusters.

\section{Discussion}

\subsection{Comparison between the two galaxies}

Although NGC~4565 and NGC~5907 have similar total luminosities and
Hubble types, there are significant differences in the disk and bulge 
properties of the two galaxies. If the properties of the disk and bulge of
a spiral galaxy influence the formation and evolution of its globular
cluster system, we would therefore expect to find differences in the
GCSs of these two galaxies.

The metal--rich globular cluster population in the Milky Way shares the 
kinematics, abundances and spatial distribution of the bulge and thick
disk (e.g.~Armandroff 1993; Burkert \& Smith 1997). In the absence of a
prominent bulge and thick disk (as in NGC 5907), we could have expected
the absence of metal--rich globular clusters and the presence of a halo
population only. The globular cluster system of NGC 5907 would have been
less flattened than the one of NGC 4565 (whereas the opposite is observed
as discussed above), and the total number of 
clusters normalized to the galaxy mass
would be lower by 20\% to 30\% in NGC 5907 compared to NGC 4565
(estimated from the fraction of metal--rich to
metal--poor clusters in the Milky Way). The latter could be the case (see
Sect.~3.3) but for the contradictory reason that NGC 5907 lacks halo
clusters and not disk clusters.

Given the uncertainties in reddening (i.e.~metallicity determination
from the colors), we cannot allocate individual clusters in these two galaxies 
to the disk/bulge or halo. This awaits spectroscopic measurements of individual
clusters.
However, the fact that the globular cluster systems of NGC 4565 and NGC 5907 are
still very similar despite difference in the thick disk and bulge 
characteristics suggests that the processes which form the bulge and
thick disk are largely unrelated to the processes which form the
total globular cluster systems around late-type galaxies. 

\subsection{Comparison with the Milky Way and other Sb--Sbc spirals}

As mentioned above,
both NGC~4565 and NGC~5907 are similar in many respects to the Milky Way.
Not only are their morphological
types similar (Sb and Sbc compared to the Sbc Milky
Way), but also their absolute magnitudes ($M_V=-21.4$ and $-21.2$
compared to $M_V=-21.3$ for the Milky Way, derived from the Tully-Fisher
relation by adopting $v_c=220$\kms) and by implication their total
stellar masses.

It is therefore notable, although perhaps not surprising, that all
three galaxies have very similar {\it numbers} of globular clusters.
Independently of the total number of Milky Way clusters, NGC 4565 has
about 20\% more clusters than the Milky Way (but is also slightly more
massive), while NGC 5907 has the same number of globular clusters as the
Milky Way within the errors. Even large errors in our adopted distance
would leave all numbers within 50\% of each other. In terms of specific
frequency, the Milky Way has $S_N=0.54\pm0.12$ for an
assumed total magnitude of $M_V=-21.3\pm0.2$ and $180\pm20$ globular
clusters, identical to the value derived for the two other spirals.

Five other spirals of type Sb to Sc have studied globular cluster
systems. Their $S_N$ values are $0.9\pm0.2$ (M31, Sb), $0.2\pm0.1$ (NGC
253, Sc), $1.7\pm0.5$ (NGC 2683, Sb), $1.2\pm0.6$ (NGC 4216, Sb), $0.9\pm0.3$ 
(NGC 5170, Sb), all taken from the compilation of Ashman \& Zepf (1998). 
The quoted errors do not take into account the possibility of systematic
errors.
The corresponding $T$ values (using the mean $M/L$ values as in
Sect.~3.2) are: $1.6\pm0.5, 0.5\pm0.3, 2.7\pm0.9,
2.2\pm1.1$, and $1.7\pm0.6$. A maximum likelihood estimation of the mean
and dispersion (including the Milky Way $T=1.2\pm0.3$, NGC 4565 and NGC 5907)
returns $<T>=1.2, \sigma=0.2$. Within the large uncertainties, galaxies of type
Sb to Sc seem to produce a similar number of globular clusters per unit
mass. 

Should NGC 5907 have a lower $S_N$ than assumed (see Sect.~3.3),
there would be a hint for Sc galaxies to have fewer globular clusters
per unit mass. An increase of the number of globular clusters per unit
mass along the Hubble sequence of late-type galaxies is strengthened by 
observations of 4 Sa and Sab galaxies (NGC 3031 (=M81), NGC 4569, NGC
4594, and NGC 7814). The respective $S_N$ values are $0.7\pm0.1$,
$1.9\pm0.6$, $2\pm1$, $3.5\pm1.1$ (taken from the compilation in Ashman
\& Zepf 1998), and the respective $T$ values are $1.5\pm0.3$,
$3.6\pm1.2$, $4.6\pm2.5$, and $6.6\pm2.3$. The mean $T$ value for these earlier
types is $<T>=3.0, \sigma=1.2$, but note that, except for M81, these
values are poorly defined (either the number of globular clusters and/or the
assumed distance is uncertain).

\section{Summary and Conclusions}

We have studied the globular cluster systems around
two edge-on spiral galaxies, NGC~4565 and NGC~5907.
For both these galaxies we derive a specific frequency of $S_N\simeq0.6$,
indistinguishable from the value of the Milky Way GCS. The similarity
of the specific frequencies is notable since the
bulge-to-disk ratio and prominence of the stellar thick disk both differ
considerably between the three galaxies.
This result suggests
that the properties of the thick disk and bulge of spiral galaxies do not have
a significant influence on the
building of globular cluster systems around such galaxies.

There is some evidence for flattening of the globular cluster system
of NGC~5907, but we are unable to determine whether this is driven by
a flattened halo population of clusters or a thick disk system. The
latter possibility seems somewhat unlikely given the upper limits on
any stellar thick disk in this galaxy. An interesting alternative is
that the globular clusters in NGC~5907 are following the profile of
a significantly flattened halo.

\acknowledgments
 
We are very grateful to our Program Coordinator, Max Mutchler, whose
work and help greatly assisted this project.
M.K.-P. gratefully acknowledges support from the Alexander von Humboldt 
Foundation.
K.M.A. and S.E.Z. acknowledge support for this project from 
NASA through the STScI grant GO-06092.01-94A.


\onecolumn


\clearpage

\begin{deluxetable}{l l l l}
\tablenum{1}
\tablecaption{
Some properties of NGC 4565 and NGC 5907
}
\tablehead{
\colhead{Property} & \colhead{NGC 4565} & \colhead{NGC 5907} & 
\colhead{Reference}\\
}

\startdata
RA (2000) & $12h 36m 20.6s$ & $15h 15m 53.8s$  & RC3 \nl
DEC(2000) & $+25^\circ 59m 05s$ & $+56^\circ 19m 46s$ & RC3 \nl
Type      &  Sb & Sc & RC3 \nl 
V$_{\rm T}^0$  & $8.58\pm0.07$ & $9.18\pm0.10$ & RC3 (value corrected
for internal and external reddening)\nl
(B--V)$_{\rm T}^0$ & $0.62\pm0.06$ & $0.52 \pm0.04$ & RC3 \nl
E(B--V) & 0.015 (+internal) & 0.011 (+internal) & Schlegel et al.~1998 \nl
Inclination & $86.5^{\circ}$ & $87^{\circ}$ & Garc\'ia-Burillo et
al.~1997, Morrison et al.~1994 \nl
Thick Disk & 5\% of B$_{\rm T}$ & $<0.6$\% of thin disk & van der 
              Kruit \& Searle 1981, Morrison et al.~1994 \nl 
Position Angle & $135.5^\circ \pm0.5$ & $156^\circ$ & Rupen et
              al.~1991, Garc\'{\i}a-Burillo et al.~1997 \nl
v$_{\rm helio}$ & $1227\pm4$\kms & $522\pm40$\kms & RC3 \nl
Adopted distance & $10$ Mpc & $12$ Mpc & Jacoby et al.~1996, Bottinelli
et al.~1988, Sch\"oniger \& Sofue 1994 \nl
$(m-M)$ & 30.0 & 30.4 & derived from the above distance \nl
$M_{V_T^0}$ & $-21.4$ & $-21.2$ & derived from the above values \nl
\enddata
\tablenotetext{}{
RC3: de Vaucouleurs et al.~1991
}
\end{deluxetable}


\clearpage

\begin{deluxetable}{l ll rr r r}
\tablenum{2}
\tablecaption{
Globular clusters in NGC 4565 
}
\tablehead{
\colhead{ID} & \colhead{RA(2000)} & \colhead{DEC(2000)} &
\colhead{Y$^a$} & \colhead{Z$^b$} & \colhead{B} & \colhead{(B--I)} \\
\colhead{} & \colhead{} & \colhead{} & \colhead{arcsec} &
\colhead{arcsec} & \colhead{mag} &  \colhead{mag}
}

\startdata
KAZF4565-1	&12 36 13.8 &$+$26 00 35.3 &-11.6 &-122.4 &$22.66\pm 0.04$ &$2.06\pm 0.05$\nl
KAZF4565-2	&12 36 14.0 &$+$25 59 58.2 &-35.8 & -94.6 &$22.88\pm 0.04$ &$1.49\pm 0.05$\nl
KAZF4565-3	&12 36 15.7 &$+$26 00 41.0 & 11.1 &-108.5 &$22.09\pm 0.02$ &$1.50\pm 0.03$\nl
KAZF4565-4	&12 36 16.5 &$+$25 58 25.0 &-76.8 &  -3.7 &$23.09\pm 0.04$ &$1.69\pm 0.06$\nl
KAZF4565-5	&12 36 17.4 &$+$25 59 16.8 &-31.6 & -32.2 &$22.37\pm 0.03$ &$1.90\pm 0.04$\nl
KAZF4565-6	&12 36 17.5 &$+$25 59 24.0 &-25.7 & -36.4 &$23.33\pm 0.05$ &$1.97\pm 0.08$\nl
KAZF4565-7	&12 36 18.0 &$+$25 58 51.2 &-44.7 &  -9.0 &$21.49\pm 0.02$ &$1.60\pm 0.02$\nl
KAZF4565-8	&12 36 18.0 &$+$25 59 26.5 &-20.0 & -34.3 &$22.20\pm 0.03$ &$1.68\pm 0.04$\nl
KAZF4565-9	&12 36 18.1 &$+$25 59 29.8 &-15.9 & -35.0 &$22.36\pm 0.03$ &$1.94\pm 0.04$\nl
KAZF4565-10	&12 36 18.4 &$+$25 59 30.5 &-12.7 & -32.8 &$22.91\pm 0.05$ &$1.70\pm 0.07$\nl
KAZF4565-11	&12 36 18.5 &$+$25 59 26.9 &-14.7 & -29.5 &$22.59\pm 0.04$ &$1.87\pm 0.05$\nl
KAZF4565-12	&12 36 18.8 &$+$25 59 04.9 &-27.4 & -11.3 &$21.04\pm 0.01$ &$1.80\pm 0.02$\nl
KAZF4565-13	&12 36 18.9 &$+$25 59 01.7 &-28.1 &  -7.7 &$23.46\pm 0.07$ &$1.50\pm 0.10$\nl
KAZF4565-14	&12 36 19.2 &$+$25 59 25.4 & -9.2 & -22.4 &$23.29\pm 0.08$ &$2.23\pm 0.11$\nl
KAZF4565-15	&12 36 19.7 &$+$26 00 04.7 & 23.2 & -45.3 &$22.78\pm 0.04$ &$1.78\pm 0.05$\nl
KAZF4565-16	&12 36 19.8 &$+$25 58 55.9 &-23.4 &   5.1 &$23.13\pm 0.06$ &$1.66\pm 0.08$\nl
KAZF4565-17	&12 36 19.9 &$+$25 59 20.4 & -5.4 & -11.3 &$22.96\pm 0.07$ &$1.64\pm 0.10$\nl
KAZF4565-18	&12 36 20.0 &$+$25 59 14.3 & -8.7 &  -6.2 &$23.35\pm 0.06$ &$2.42\pm 0.08$\nl
KAZF4565-19	&12 36 20.2 &$+$25 59 56.0 & 22.0 & -34.8 &$23.45\pm 0.06$ &$2.00\pm 0.08$\nl
KAZF4565-20	&12 36 20.3 &$+$25 59 06.4 &-11.5 &   2.0 &$22.49\pm 0.04$ &$2.06\pm 0.06$\nl
KAZF4565-21	&12 36 20.8 &$+$25 58 31.8 &-30.8 &  31.8 &$22.64\pm 0.03$ &$2.05\pm 0.05$\nl
KAZF4565-22	&12 36 21.1 &$+$25 59 36.6 & 17.8 & -11.3 &$23.10\pm 0.05$ &$2.75\pm 0.07$\nl
KAZF4565-23	&12 36 21.2 &$+$25 58 34.7 &-25.3 &  33.2 &$23.32\pm 0.06$ &$1.84\pm 0.08$\nl
KAZF4565-24	&12 36 21.6 &$+$25 59 44.2 & 27.3 & -12.8 &$23.21\pm 0.05$ &$2.23\pm 0.08$\nl
KAZF4565-25	&12 36 21.9 &$+$25 58 18.1 &-29.6 &  52.2 &$23.02\pm 0.05$ &$1.56\pm 0.07$\nl
KAZF4565-26	&12 36 22.3 &$+$25 58 30.4 &-17.6 &  46.6 &$23.04\pm 0.05$ &$1.68\pm 0.07$\nl
KAZF4565-27	&12 36 22.4 &$+$25 58 40.8 & -8.8 &  40.6 &$23.38\pm 0.07$ &$2.39\pm 0.10$\nl
KAZF4565-28	&12 36 22.8 &$+$25 58 01.6 &-32.9 &  71.8 &$22.29\pm 0.03$ &$1.66\pm 0.04$\nl
KAZF4565-29	&12 36 23.4 &$+$25 59 10.0 & 20.5 &  28.7 &$22.22\pm 0.03$ &$2.21\pm 0.04$\nl
KAZF4565-30	&12 36 23.7 &$+$25 59 10.0 & 24.0 &  32.0 &$22.90\pm 0.04$ &$2.65\pm 0.06$\nl
KAZF4565-31	&12 36 24.3 &$+$25 59 08.2 & 28.1 &  38.6 &$22.03\pm 0.02$ &$2.05\pm 0.04$\nl
KAZF4565-32	&12 36 25.4 &$+$25 58 21.7 &  6.2 &  82.0 &$22.74\pm 0.04$ &$1.20\pm 0.06$\nl
KAZF4565-33	&12 36 25.4 &$+$25 58 33.2 & 14.1 &  74.0 &$23.00\pm 0.04$ &$2.46\pm 0.06$\nl
KAZF4565-34	&12 36 25.5 &$+$25 58 36.1 & 17.3 &  72.9 &$23.45\pm 0.06$ &$2.21\pm 0.09$\nl
KAZF4565-35	&12 36 25.9 &$+$25 58 47.6 & 28.7 &  68.1 &$23.15\pm 0.05$ &$1.81\pm 0.07$\nl
KAZF4565-36	&12 36 26.1 &$+$25 57 26.6 &-25.9 & 127.8 &$22.49\pm 0.03$ &$1.66\pm 0.05$\nl
KAZF4565-37	&12 36 26.1 &$+$25 58 42.2 & 27.3 &  74.1 &$23.42\pm 0.06$ &$2.58\pm 0.08$\nl
KAZF4565-38	&12 36 26.6 &$+$25 58 08.4 &  8.7 & 103.2 &$23.32\pm 0.06$ &$1.57\pm 0.08$\nl
KAZF4565-39	&12 36 26.9 &$+$25 58 00.8 &  6.1 & 111.2 &$23.13\pm 0.05$ &$1.25\pm 0.07$\nl
KAZF4565-40	&12 36 27.4 &$+$25 58 16.0 & 21.5 & 105.5 &$23.21\pm 0.05$ &$1.97\pm 0.07$\nl
\enddata

\tablenotetext{a}{Y: Distance from the center along the disk }
\tablenotetext{b}{Z: Distance orthogonal from the disk }
\tablenotetext{}{
}
\end{deluxetable}

\clearpage

\begin{deluxetable}{l ll rr r r}
\tablenum{3}
\tablecaption{
Globular clusters in NGC 5907 
}
\tablehead{
\colhead{ID} & \colhead{RA(2000)} & \colhead{DEC(2000)} &
\colhead{Y$^a$} & \colhead{Z$^b$} & \colhead{B} & \colhead{(B--I)} \\
\colhead{} & \colhead{} & \colhead{} & \colhead{arcsec} &
\colhead{arcsec} & \colhead{mag} &  \colhead{mag}
}

\startdata

KAZF5907-1	& 15 15 40.1 & $+$56 20 03.8 & -96.7 &  -64.6 & $22.23\pm 0.02$ & $1.45\pm 0.04$\nl
KAZF5907-2	& 15 15 46.7 & $+$56 19 50.9 & -51.8 &  -29.7 & $21.89\pm 0.02$ & $1.82\pm 0.03$\nl
KAZF5907-3	& 15 15 46.9 & $+$56 21 14.8 & -15.2 & -105.1 & $22.48\pm 0.03$ & $1.76\pm 0.04$\nl
KAZF5907-4	& 15 15 47.3 & $+$56 21 31.3 &  -5.6 & -118.8 & $23.34\pm 0.07$ & $1.53\pm 0.10$\nl
KAZF5907-5	& 15 15 47.7 & $+$56 20 52.8 & -18.5 &  -82.4 & $23.11\pm 0.05$ & $1.50\pm 0.07$\nl
KAZF5907-6	& 15 15 49.1 & $+$56 20 28.0 & -18.3 &  -55.0 & $23.02\pm 0.05$ & $1.61\pm 0.07$\nl
KAZF5907-7	& 15 15 50.0 & $+$56 20 03.1 & -22.3 &  -29.4 & $23.15\pm 0.05$ & $1.28\pm 0.08$\nl
KAZF5907-8	& 15 15 51.5 & $+$56 19 28.6 & -25.5 &    7.4 & $22.68\pm 0.04$ & $2.13\pm 0.05$\nl
KAZF5907-9	& 15 15 51.8 & $+$56 19 49.1 & -14.2 &  -10.1 & $23.28\pm 0.06$ & $1.32\pm 0.09$\nl
KAZF5907-10	& 15 15 52.2 & $+$56 20 38.8 &   9.8 &  -53.6 & $23.30\pm 0.06$ & $1.77\pm 0.08$\nl
KAZF5907-11	& 15 15 52.5 & $+$56 20 11.4 &  -0.1 &  -28.0 & $23.38\pm 0.07$ & $1.90\pm 0.10$\nl
KAZF5907-12	& 15 15 53.1 & $+$56 19 01.2 & -25.0 &   37.8 & $22.95\pm 0.03$ & $2.53\pm 0.05$\nl
KAZF5907-13	& 15 15 53.4 & $+$56 19 22.1 & -13.7 &   20.0 & $23.26\pm 0.05$ & $1.20\pm 0.08$\nl
KAZF5907-14	& 15 15 54.2 & $+$56 20 00.2 &   8.6 &  -11.6 & $22.68\pm 0.04$ & $1.97\pm 0.05$\nl
KAZF5907-15	& 15 15 54.5 & $+$56 20 02.4 &  11.0 &  -12.8 & $22.64\pm 0.04$ & $1.92\pm 0.05$\nl
KAZF5907-16	& 15 15 54.8 & $+$56 19 39.4 &   4.0 &    9.1 & $23.27\pm 0.05$ & $1.84\pm 0.08$\nl
KAZF5907-17	& 15 15 55.4 & $+$56 18 45.3 & -14.7 &   60.2 & $23.34\pm 0.05$ & $1.21\pm 0.07$\nl
KAZF5907-18	& 15 15 55.7 & $+$56 19 44.4 &  12.6 &    7.8 & $22.87\pm 0.04$ & $1.76\pm 0.06$\nl
KAZF5907-19	& 15 15 55.8 & $+$56 19 49.4 &  15.7 &    3.5 & $21.81\pm 0.02$ & $1.78\pm 0.03$\nl
KAZF5907-20	& 15 15 55.9 & $+$56 19 28.9 &   8.0 &   22.6 & $23.19\pm 0.04$ & $2.32\pm 0.06$\nl
KAZF5907-21	& 15 15 56.1 & $+$56 19 57.3 &  21.2 &   -2.8 & $22.46\pm 0.03$ & $1.73\pm 0.04$\nl
KAZF5907-22	& 15 15 57.3 & $+$56 19 25.3 &  17.2 &   30.9 & $23.43\pm 0.05$ & $1.72\pm 0.07$\nl
KAZF5907-23	& 15 15 57.4 & $+$56 17 47.4 & -24.1 &  119.5 & $23.30\pm 0.05$ & $1.45\pm 0.07$\nl
KAZF5907-24	& 15 15 57.6 & $+$56 19 10.9 &  13.4 &   45.1 & $23.35\pm 0.05$ & $1.21\pm 0.07$\nl
KAZF5907-25	& 15 15 57.8 & $+$56 18 12.6 &  -9.8 &   98.6 & $23.33\pm 0.05$ & $1.35\pm 0.07$\nl
\enddata 
\tablenotetext{a}{Y: Distance from the center along the disk }
\tablenotetext{b}{Z: Distance orthogonal from the disk }
\tablenotetext{}{
}
\end{deluxetable}


\clearpage

\begin{deluxetable}{l c c}
\tablenum{4}
\tablecaption{
Numbers of Milky Way globular clusters in mask
}
\tablehead{
\colhead{} & \colhead{NGC 4565 mask} & \colhead{NGC 5907 mask} \\
}
\startdata

$+Y+Z$ orientation & 29 & 24 \\
$-Y+Z$ orientation & 34 & 24 \\
$-Y-Z$ orientation & 36 & 26 \\
$+Y-Z$ orientation & 35 & 23 \\
Mean , Dispersion & $33.50\pm3.11$ & $24.25\pm1.26$ \\
 & & \\
20\% closer & $39.00\pm2.94$ & $37.75\pm2.50$ \\
20\% further & $24.25\pm1.26$ & $20.75\pm1.71$ \\

\enddata 
\tablenotetext{}{X and Y are Galactic coordinates. The results for the
closer and further sample are the means of all 4 orientations.
}
\end{deluxetable}


\clearpage

\begin{figure}
\psfig{figure=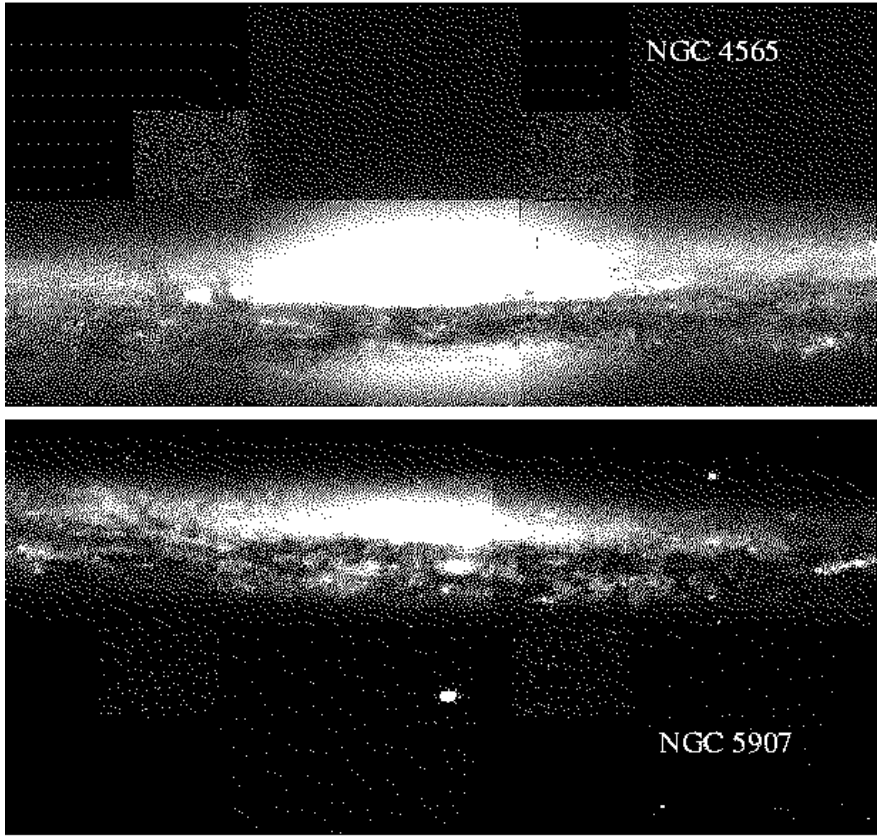,height=16cm,width=16cm
,bbllx=8mm,bblly=8mm,bburx=205mm,bbury=205mm}
\caption{
Mosaic of the two WFPC2 pointings on our target galaxies NGC 4565 and NGC
5907. 
}
\end{figure}

\clearpage

\begin{figure}
\psfig{figure=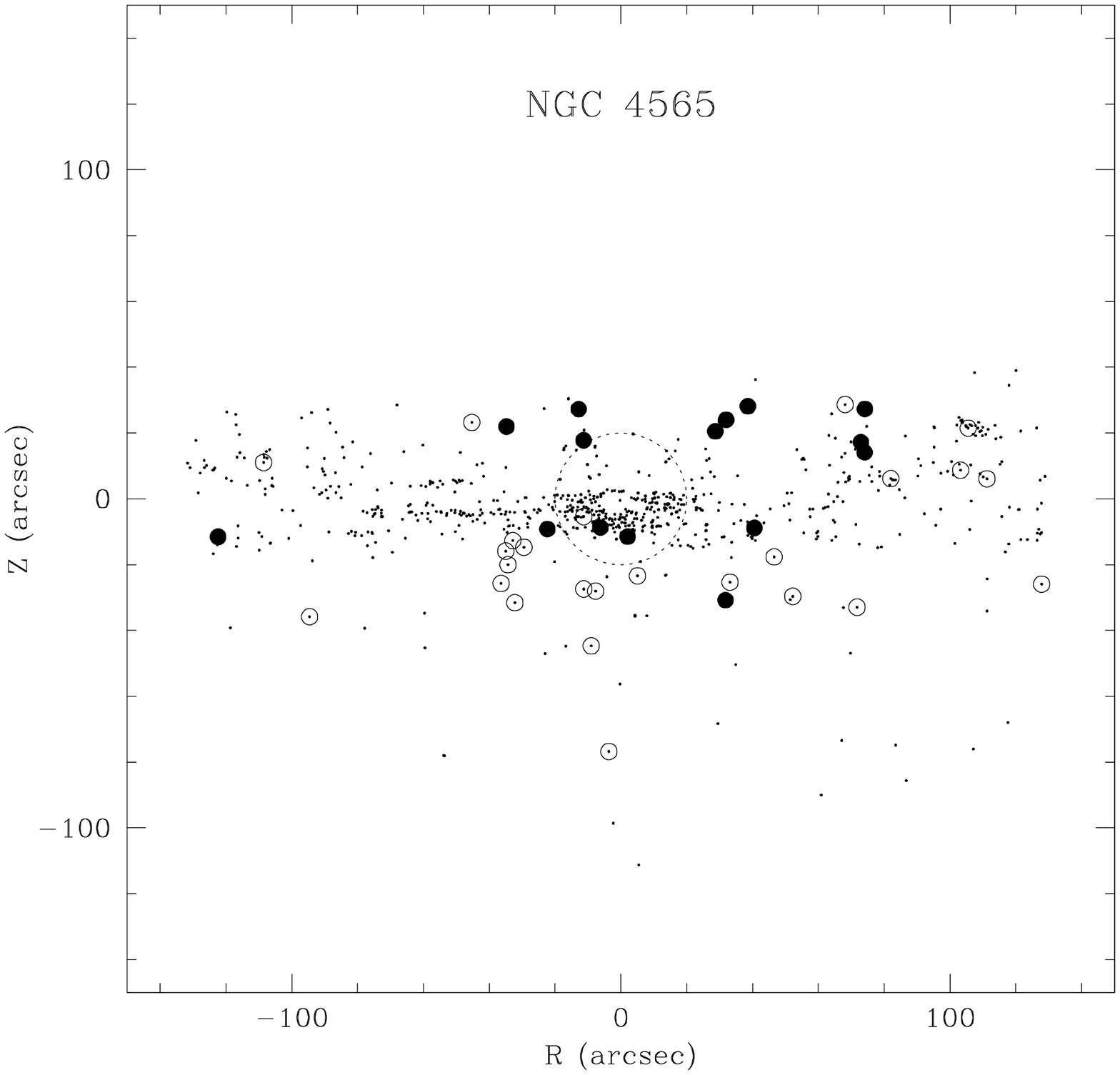,height=8cm,width=8cm
,bbllx=8mm,bblly=57mm,bburx=205mm,bbury=245mm} 
\psfig{figure=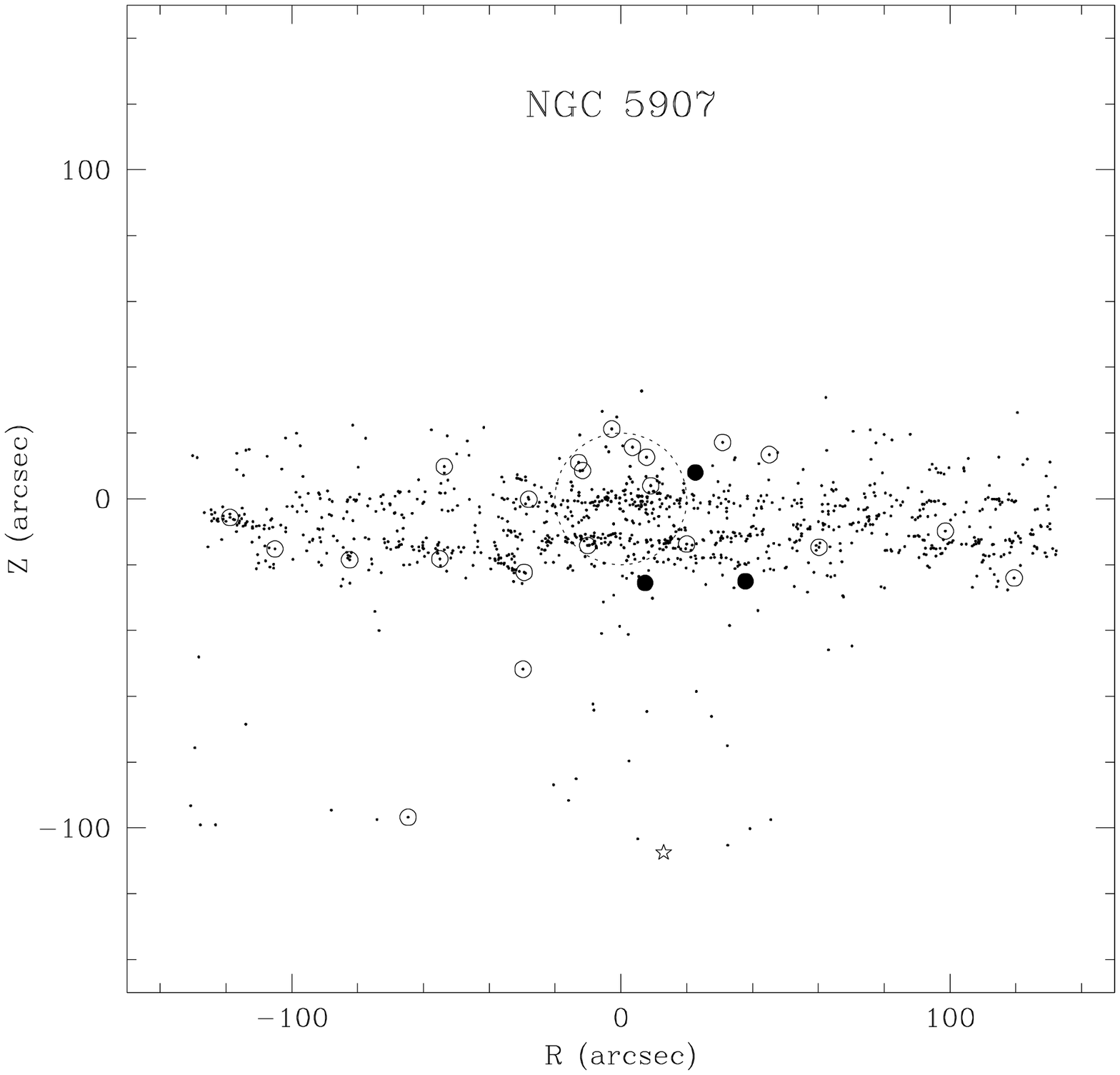,height=8cm,width=8cm
,bbllx=8mm,bblly=57mm,bburx=205mm,bbury=245mm}
\caption{
Position of all detected objects around NGC 4565 and NGC 5907. 
Circles mark the position
of the globular clusters: open symbols are used for clusters with
B--I$<2.0$, filled symbols mark redder objects. The latter are redder than the 
reddest Milky Way globular cluster (after de-reddening), i.e.~these objects
that are almost certainly affected by dust.
The star in NGC 5907 marks an object that passed the selection criteria
for globular clusters, but was identified as a star visually.
}
\end {figure}

\clearpage

\begin{figure}
\psfig{figure=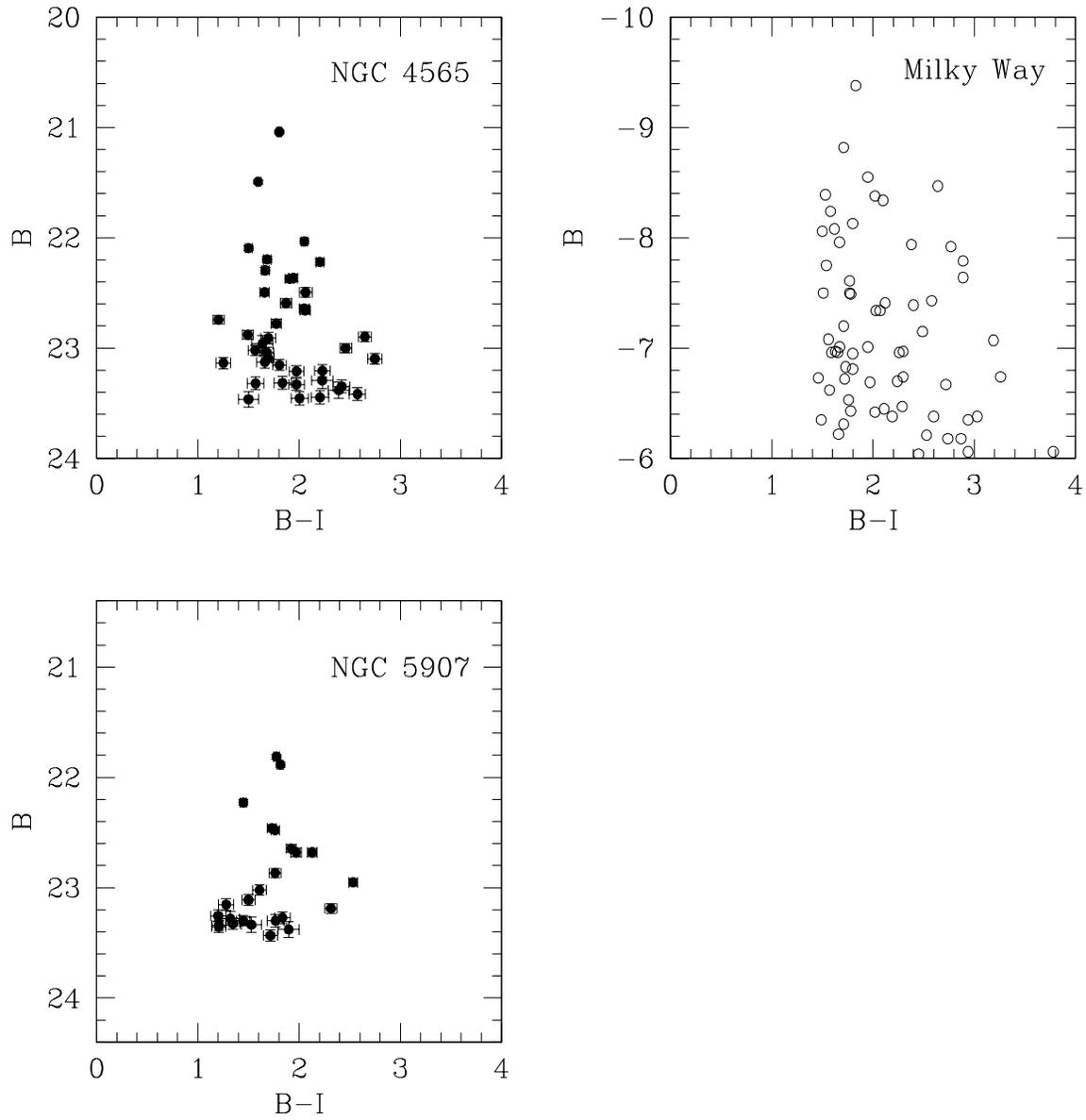,height=16cm,width=16cm
,bbllx=8mm,bblly=57mm,bburx=205mm,bbury=245mm}
\caption{
Color magnitude diagrams for globular clusters in NGC 4565, NGC 5907 and
the Milky Way. Note that the Milky Way sample was neither corrected for
reddening, nor truncated spatially to match the samples of the two other
galaxies. The scales were adjusted as to roughly show clusters one
magnitude brighter than the turn-over magnitude.
}
\end {figure}

\clearpage

\begin{figure}
\psfig{figure=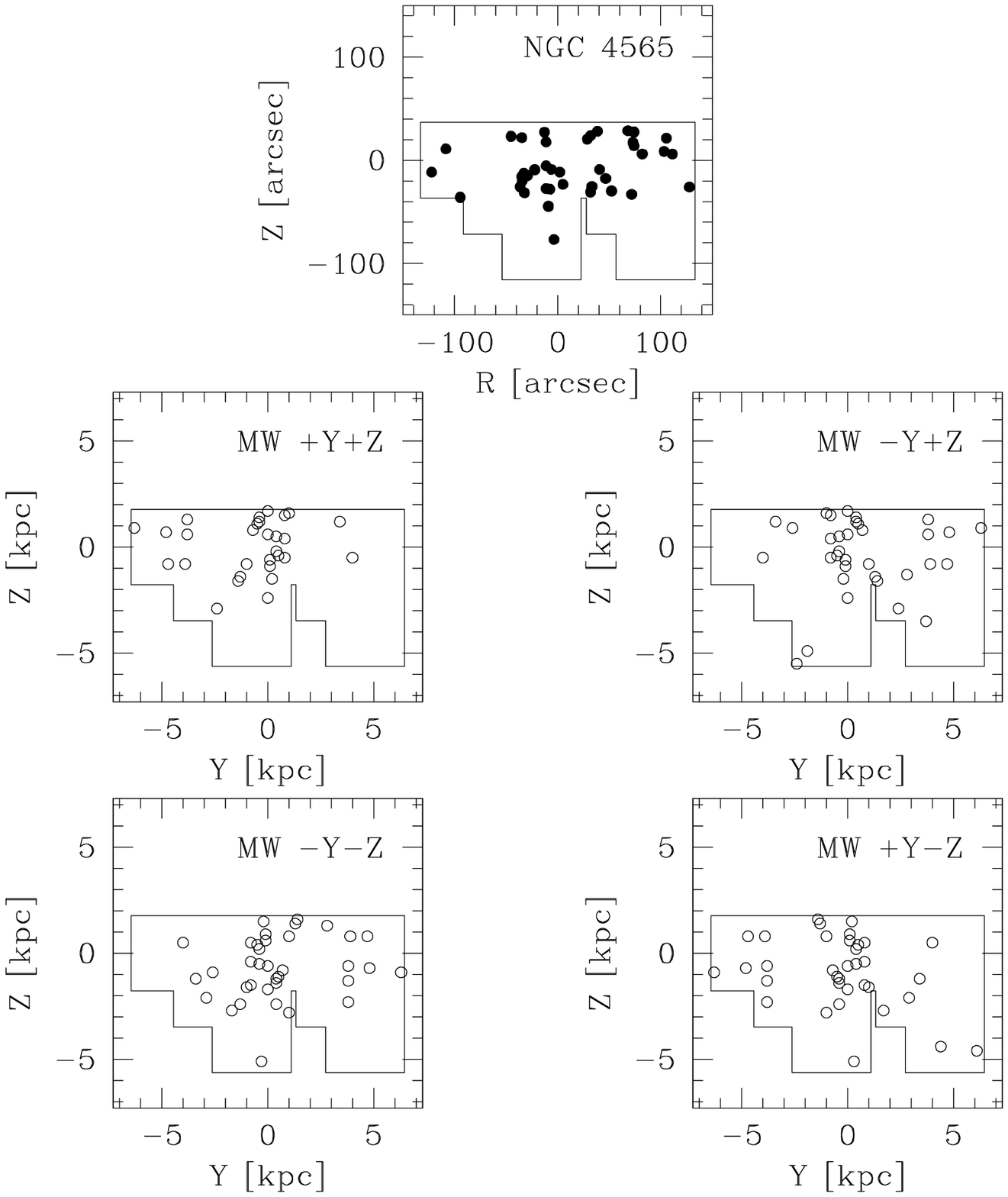,height=16cm,width=16cm
,bbllx=8mm,bblly=57mm,bburx=205mm,bbury=245mm} 
\caption{Milky Way clusters in the NGC 4565 mask (see text for details).
}
\end{figure}

\clearpage

\begin{figure}
\psfig{figure=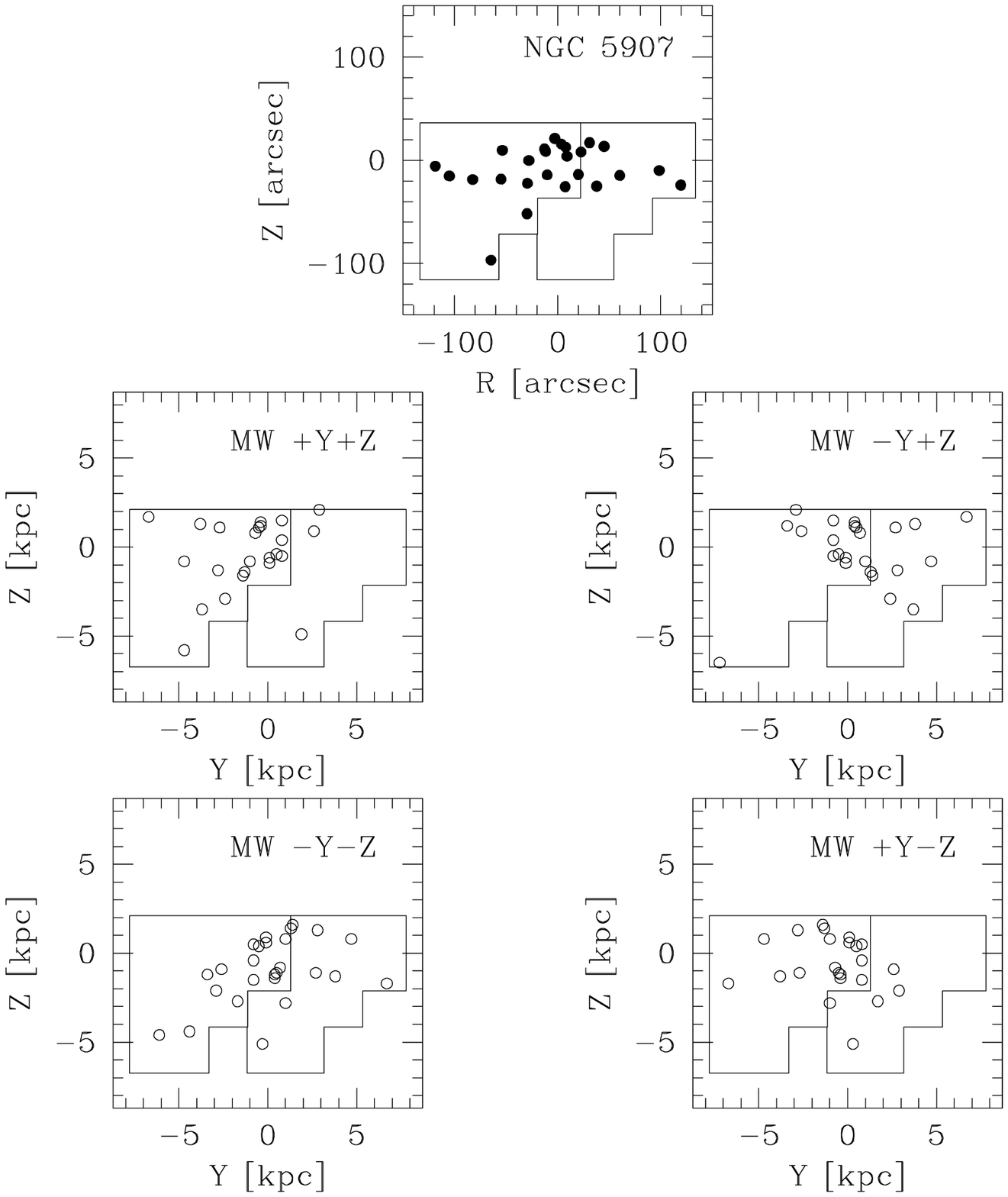,height=16cm,width=16cm
,bbllx=8mm,bblly=57mm,bburx=205mm,bbury=245mm} 
\caption{Milky Way clusters in the NGC 5907 mask (see text for details).
}
\end{figure}


\end{document}